# Diversity of anisotropy effects in the breakup of metallic FCC nanowires into ordered nanodroplet chains

Vyacheslav N. Gorshkov [1], Vladimir V. Tereshchuk [1], and Pooya Sareh [2*]

[1]National Technical University of Ukraine, Igor Sikorsky Kyiv Polytechnic Institute, 37 Prospect Peremogy, Kiev 03056, Ukraine.

[2]School of Engineering, University of Liverpool, London Campus, 33 Finsbury Square, London EC2A 1AG, United Kingdom.
*Corresponding author. Email: pooya.sareh@liverpool.ac.uk

Keywords: face-centered cubic crystal structure, nanowire breakup, nanostructure instabilities, anisotropy of surface energy density, Monte-Carlo kinetic model

**Abstract**

We have analyzed the expressed manifestation of the anisotropy of surface energy density in the dynamics of ultrathin nanowires, which break up into disjointed clusters when annealed below their melting temperature. The breakup process is studied for different temperatures and orientations of the nanowire axis relative to its internal crystal structure using the Monte Carlo kinetic method. We have also presented an approximate analytical model of the instability of nanowires. Generally, the interpretation of experimental results refers to the theoretical model developed by Nichols and Mullins, which is based on conceptions about the Rayleigh instability of liquid jets. In both cases, the theories - which do not take into account the anisotropy of surface energy density - predict the breakup of a nanowire/liquid jet with radius $r$ into fragments with an average length $\Lambda = 9r$. However, the observed value, $\Lambda/r$, often deviates from 9 either to lower values or to substantially greater ones (up to 24-30). Our results explain various observed features of the breakup and the significant variations in the values of its parameter $\Lambda/r$ depending on experimental conditions. In particular, the ambiguous role of exchange by atoms of the surface of a nanowire with the surrounding layer of free atoms formed as a result of their rather intensive sublimation, which occurs in a number of cases, has been investigated. We have shown that this exchange can lead both to a decrease, and to a significant increase, in the parameter $\Lambda/r$. The obtained results could be potentially useful in applications such as the development of optical waveguides based on ordered nanoparticles chains.

# 1. Introduction

Nowadays, metallic and semiconductor nanowires draw considerable attention due to their beneficial physical and chemical properties. Specifically, mechanical and electrical properties of gold nanowires have led to their exploitation for bioelectrical signal detection **[1-3]**. At the same time the possibility of tuning the band gap width in silicon nanowires makes them particularly suitable for optoelectronics applications **[4-5]**. Moreover, modern methods of nanowire synthesis **[6-11]** make it possible to tune the diameter-modulation periodicity and cross-sectional anisotropy of nanowires, which allows the fabrication of 1D-structures with different surface morphologies. For example, the large stretchability of silicon nanowire springs **[12]** - produced by using low-temperature thin-film technology - makes them ideal candidates for the role of building blocks of bio/mechanical sensors **[13-17]**. However, due to the large surface to volume ratio, nanowires demonstrate poor thermal stability and breakup into isomeric nanoparticles at premelting temperatures **[18-21]**. On the one hand, this phenomenon can significantly impair optoelectrical properties of the devices. On the other hand, nanowire instability can be utilized in the fabrication of long chains of isomeric nanoparticles which can be used for the construction of plasmon waveguides capable of transporting electromagnetic energy below the diffraction limit **[22, 23]**.

The existing analytical model of the nanowires breakup by Nichols and Mullins **[24]** is very close to the model of Plateau-Rayleigh instability in liquid inviscid threads **[25]** because it is constructed under the assumption that the surface energy density, $\sigma$, of the nanowire is isotropic. Despite the fact that internal flows are absent in the case of nanowire breakup and, according to the statement in **[24]**, only the surface diffusion of atoms is responsible for nanowire dynamics, the wavelength of periodic perturbation of its radius with maximal growth increment, $\lambda_{max}$, is the same as in cylindrical liquid streams, i.e. $\lambda_{max} \approx 9r_0$, where $r_0$ is the initial radius of the nanowire. The origin of this coincidence is that, in both cases, changes in total surface energy and the displacement of center of mass contained in a half-wavelength segment (between necking and widening) for the supposed sinusoidal modulations of radius are similar.

The correspondence of the predictions of the theoretical model **[24]** with the experimentally measured average sizes of nanodroplets and the distances between them, $\Lambda$, was noted in a number of previous studies performed for Cu, Ag, Au, Pt-nanowires **[26-30]**. The discrepancies in the experimental and theoretical values, which are approximately 10-16% as noted by the authors, are quite expected, taking into account the assumptions of the model itself and the measurement errors in the experiments as well as the role of uncontrolled factors. However, there are also known cases of extremely large values of the breakup parameter, $\Lambda/r_0 \sim 25 - 30$ **[31]**. Moreover, significant periodic modulations of the nanowire radius ($\Delta r_{max} \sim 0.5 r_0$) can be observed in a high-resolution TEM photo **[32]** with a wavelength of $\lambda_{max} \approx 5.7 r_0$, which is below the classical threshold for the development of instability **[24]**, $\lambda_{max} > 2\pi r_0$. Therefore, the physical mechanisms responsible for such dispersion in the parameters of this instability require careful future research.

The observed deviations in the average distance between the centers of neighbor nanodroplets, $\Lambda$, which was associated with the wavelength $\lambda_{max}$, must be interpreted as a result of the anisotropy of surface energy density. The first discussion of this factor [33, 34] resulted in a clear qualitative estimation: if $\sigma$ decreases/increases at the necking regions relative to the widening regions when the surface perturbations arise, then this decrease/increase leads to acceleration/inhibition in the development of the instability. Less obvious and more complex is the answer to the question how $\sigma$ inhomogeneity affects the value $\Lambda$, and why in some experiments a strong dependency of $\Lambda$ on temperature, $T$, is observed [26, 28]. This is the fact that the well-known model [24] cannot explain, as well as the processes that are responsible for giant wavelength perturbations, $\Lambda \sim 30 r_0$, as reported in mentioned studies [28, 31] (see Fig. 1(a)). Finally, we would like to note a series of publications [35-38] in which the physical processes concerning the nanowire dynamics were studied in the frameworks of stochastic atomistic modeling and molecular dynamics. However, the physical mechanisms determining the experimentally observed diversity of the breakup parameters, $\Lambda$ and $\lambda_{max}$, (in general, the origin of these values is different [39-40]) are not sufficiently clear in detail.

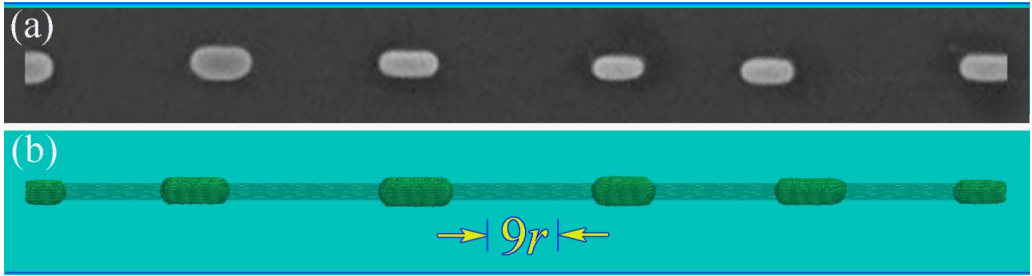

**Fig. 1.** (a) Experimental configuration obtained for an Au-nanowire at temperature $T = 400°$ C adapted from Vigonski *et al.* [31]. (b) Result of our numerical experiments with a nanowire ($L/r = 124$ and $\Lambda/r \approx 25$) oriented along the [110]-direction; the initial nanowire shape is also outlined.

The effects associated with $\sigma(r)$ anisotropy are expected to manifest when the axis of the nanowire does not coincide with the axis of symmetry of the crystal lattice. That is why these 'asymmetrical' orientations were studied in our previous work [39]. But it turned out, as shown in our present work, that the most striking deviations in the dynamics of the nanowire, in comparison with model [24], are manifested precisely in the axisymmetric orientations [100], [111], and [110] (the nanowire axis is the axis of symmetry of the fourth-, third-, and second-order, respectively). A variety of scenarios in terms of the physical factors determining the breakup parameter, $\Lambda$, is based on that in these orientations the nanowire lateral surface is mainly composed of faces of (100), (111), and (110) types with the lowest surface energy densities, $\sigma(r)$. Therefore, the formation of the nanowire constrictions is accompanied by the appearance of a noticeable imbalance between the values, $\sigma(r)$, at the areas of narrowing and bulging of the nanowire in one direction or another, the total free-energy decreasing in any case.

It is to be noted that, in addition to the surface diffusion of atoms, detachment-reattachment processes can play an important role in the development of nanowire instabilities. We have established that the role of the exchange of atoms strongly depends on the orientation of a nanowire and can lead either to a significant reduction in the wavelengths of excited surface perturbations or to their considerable increase (up to $\lambda_{max} \approx (25 - 30) \times r_0$; see Fig. 1(b) and Section 4), as compared with the results obtained in cases in which the detachment processes are blocked. Also, the approximate analytical model of the disintegration process is given taking into account the anisotropy of the surface energy density (see Section 3), the results of which predict the effects observed in the experiments.

All the results of our numerical experiments were interpreted on the basis of clear qualitative physical concepts that represent in details the features of breakup scenarios, which can be realized in the FCC-nanowires dynamics depending on experimental conditions. The exploited Monte Carlo method, operating with two dimensionless parameters, has been successfully used earlier in a series of articles dedicated to modeling the surface- and nanotube-templated growth **[41, 42]** of nanoparticles and other nanostructures, as well as the sintering of the former **[43, 44]**. The detailed description of this method can be found in our previous study **[39]**. However, its main ideas will be briefly outlined below (see Section 2).

## 2. Main concepts of the exploited Monte-Carlo model

Atoms that constitute a nanowire locate at FCC lattice sites with coordination number $m_c = 12$. Every surface/movable atom, which interacts with $m_0 < m_c$ nearest neighbors, can randomly hop to any nearest adjoining vacancies. Probability of such hopping to occur is proportional to $\exp(-m_0\Delta/kT) \equiv p^{m_0}$, where $m_0\Delta > 0$ presents the activation free-energy barrier, $p = \exp(-\Delta/kT)$, and $T$ is temperature. If the hop is to be realized, the candidates for the target site are $n_t = (m_c - m_0 + 1)$ sites: the original site and $n = m_c - m_0$ of the nearest vacancies in which the numbers of nearest occupied sites are equal to $m_t^{(i)}$ ($i = 1,2,3,\ldots,n$). The probability of each supposed hopping is proportional to temperature-dependent Boltzmann factors $\exp(m_t|\varepsilon|/kT)$ normalized over $n_t$ possible positions ($m_t \in \{m_0, m_t^{(i)}\}$ is the number of nearest neighbors at the target sites). The value $\varepsilon < 0$ represents the pair binding free-energy. Note that $m_t = 0$ at the target site corresponds to detachment for the bound atom when it rejoins the surrounding 'gas' of free atoms.

The free atoms diffuse by hopping with random angles of scattering and fixed length, $\ell$, of the hops ($\ell = a/\sqrt{2}$, the lattice spacing, $a$, is taken as the unit of length in sequel). They can be reattached to the nanowire surface (at unoccupied lattice sites that are the nearest-neighbors to the nanostructure atoms) if these hop into a cube with edge length $a/2$, centered at such a site. The nanowire axis is also the axis of a cylindrical container with walls that reflect the free atoms. The radius of this container is much greater than the nanowire radius.

In order to evaluate the role of detachment-reattachment processes in the nanowire break-up, we also conducted numerical experiments in which the evaporation from the nanowire surface was reduced. For this, an additional coefficient $P_{filter} < 1$ sometimes was used in the model to decrease the detachment probability of bound atoms by $1/P_{filter}$ times.

To avoid the shrinkage of the finitely-long nanowire and early generation of nanodroplets at its ends because of the so-called "end-effect" **[39, 40]**, we consider long nanowires with 'frozen' ends. In this case, 5-7 atomic planes of moveless/frozen atoms, which are oriented perpendicular to the nanowire axis, connect the nanowire butt-ends with the end-walls of the container. This approach is an analogue of the periodic boundary condition.

In accordance with prior studies of the morphology dynamics of FCC nanoclusters **[39-47]**, the present mesoscale model operates by two dimensionless thermodynamic parameters,

$$\alpha = |\varepsilon|/kT \quad \text{and} \quad p = \exp(-\Delta/kT), \tag{1}$$

with reference 'intermediate temperature' values $\alpha_0 = 1$ and $p_0 = 0.7$, which establish the link between the values of $p$ and $\alpha$ when temperature $T$ changes, expressed as

$$p = p_0^{\alpha/\alpha_0}. \tag{2}$$

Thus, the general patterns of nanowire breakup are examined in this article without reference to any specific metals.

## 3. Effect of the anisotropy of surface energy density in nanowires breakup

The wavelength of surface perturbations with maximum increment is estimated by the exploitation of the generalized equation of motion for a substance enclosed in a half-wave segment of a nanowire/liquid jet (in a region between the minimum and maximum radius) with mass $m_{\lambda/2}$ ($m_{\lambda/2} \sim \lambda$). The increase in the amplitude of the initial perturbations of the radius, $\Delta r = -\varepsilon(t)\cos(2\pi x/\lambda)$, is followed by a change in surface energy, $\Delta E_{surf}$, of the analyzed segment and the displacement of its center of mass, $\Delta_{cm}$ ($\Delta_{cm} \sim \varepsilon(t) \times \lambda$). The basic equation is the well-known relation for dissipative systems

$$m_{\lambda/2}\left(\frac{d^2}{dt^2}\Delta_{cm} + \chi \frac{d}{dt}\Delta_{cm}\right) = -\frac{\partial \Delta E_{surf}}{\partial \Delta_{cm}}. \tag{3}$$

In the case of inviscid liquid jet the coefficient of dissipation, $\chi$, is equal to zero ($\chi = 0$). In the dynamics of a nanowire, the inertial term, $\frac{d^2}{dt^2}\Delta_{cm}$, can be neglected.

The calculation of $\Delta E_{surf}$ is performed for a cylinder with a variable radius along its axis, $X$, as follows

$$r(x, t=0) = r_0 - \left(\varepsilon^2/4r_0\right) - \varepsilon\cos(kx), \quad k = 2\pi/\lambda, \quad \varepsilon \ll r_0, \tag{4}$$

Eq. (3) corresponds to the conservation of nanowire/jet volume with an accuracy of $\varepsilon^2$. We assume that the surface energy density, $\sigma$, depends on an angle, $\gamma$, between the normal to the axis of the nanowire and the normal to the surface at a given point. In this approximation, the function $\sigma(\gamma)$ is even (in the cases of 'symmetrical' nanowire orientations under consideration), and for small perturbations, $\varepsilon(t)$, it has the following form

$$\sigma = \sigma_0 + \frac{1}{2}\frac{d^2\sigma}{d\gamma^2} \times \gamma^2 \approx \sigma_0(1 + \beta \times tan^2\gamma) = \sigma_0\left[1 + \beta \times \left(\frac{dr}{dx}\right)^2\right] =$$

$$= \sigma_0[1 + \beta \times (\varepsilon(t)k \times sinkx)^2], \tag{5}$$

where $\beta = \frac{1}{2}\frac{d^2}{d\gamma^2}\left(\frac{\sigma}{\sigma_0}\right)$. Then,

$$\Delta E_{surf} \sim \int_0^{\lambda/2} r\sqrt{1 + (\varepsilon(t)k \times sinkx)^2}\,[1 + \beta \times (\varepsilon(t)k \times sinkx)^2]dx - r_0\lambda/2. \quad (6)$$

With an accuracy to $\varepsilon^2$

$$\Delta E_{surf} \sim \varepsilon^2[\hat{k}^2(1 + 2\beta) - 1]/\hat{k}, \quad \hat{k} = 2\pi r_0/\lambda. \quad (7)$$

Taking into account that $m_{\lambda/2} \sim \lambda \sim 1/\hat{k}$, $\varepsilon(t) \sim \Delta_{cm}/\lambda \sim \Delta_{cm}\hat{k}$, and $\frac{\partial \Delta E_{surf}}{\partial \Delta_{cm}} = \frac{\partial \Delta E_{surf}}{\partial \varepsilon}\frac{\partial \varepsilon}{\partial \Delta_{cm}}$, Eq. (3) leads to the dispersion equations

$$\ddot{\varepsilon}(t)/\hat{k}^2 \sim \varepsilon(1 - \hat{k}^2) - \text{for an inviscid liquid jet } (\beta = 0), \quad (8a)$$

$$\chi \times \dot{\varepsilon}(t)/\hat{k}^2 \sim \varepsilon[1 - \hat{k}^2(1 + 2\beta)] - \text{for a nanowire.} \quad (8b)$$

Thus, the growth increment, $\gamma$, ($\varepsilon(t) = \varepsilon_0 \exp(\gamma t)$) reaches its maximal value, $\gamma_{max}$, for the corresponding wave vector $k_{max} = 1/\sqrt{2(1 + 2\beta)}$ in both cases under consideration, i.e.

$$\lambda_{max}/r_0 \approx 8.89 \times \sqrt{1 + 2\beta}. \quad (9)$$

In the case when the surface energy density is isotropic, $\beta = 0$, Eq. (9) gives a well-known result (obtained by Nichols and Mullins **[24]**)

If we take the surface energy density on a facet of (111) type equal to one, i.e. $\sigma_{(111)}=1$, then according to the broken-bond rule for the surface energies of noble metals [**45**], we will have $\sigma_{(100)} = 1.155$ and $\sigma_{(110)} = 1.225$. The results presented in [**31**] for Au give slightly different values for the surface energy density, but in all cases the relation $\sigma_{(111)} < \sigma_{(100)} < \sigma_{(110)}$ holds, which allows us to estimate the value of $\lambda_{Max}/r_0$ for nanowires with different orientations on the basis of Eq. (9). Such an estimation enables one to predict only the deviations of the parameter $\lambda_{Max}/r_0$ from the classical value either in a positive or negative direction. As will be shown in the next section, in some cases these deviations can significantly exceed 9 (in agreement with experiments [**28, 30, 31**]), because the presented model does not take into account some important physical factors.

## 4. Results

Here we present the results of numerical experiments with nanowires of different orientations and at different temperatures. Three temperature regimes are considered:

$$\begin{aligned}\text{«cold»} &- \alpha = 0.96 - 1.2, \; p = 0.71 - 0.65; \\ \text{«warm»} &- \quad \alpha = 0.9, \; p = 0.725; \\ \text{«hot»} &- \quad \alpha = 0.8, \; p = 0.752.\end{aligned} \quad (10)$$

The selected parameters are based on the reference values, $\alpha_0 = 1$, $p_0 = 0.7$, and relation (2). Variations of the alpha values correspond to temperature variations in numerous previous experiments ($T \approx 300° - 700°$ C). The wire diameter is $d_0 = 10a - 12.5a$ ($a$ is the FCC lattice spacing). For gold, this corresponds to $d_0 \sim 4 - 5\ nm$. The nanowire length is $L = 60r_0 - 125r_0$.

We begin our analysis of the numerical results with the most interesting case in which the wire is oriented along the [110] axis. As noted in many previous works (see, e.g., **[28, 30, 31]**), in this condition the nanowire is the most resistant to breakup and is characterized by an anomalously large value of the parameter $\Lambda/r_0 \sim 20 - 30$. The fact that the wavelength of perturbations, $\lambda_{max}$, of the nanowire radius should exceed $9r_0$ immediately before the start of breakup follows from Eq. (9) and is based on the considerations explained below.

It is known that the surface of a cylindrical nanowire is rearranged by minimizing free energy at the initial stages of evolution. Fig. 2 illustrates the structure of the lateral surface of some nanowires as a result of such transformation. In the case of the [110] orientation (Fig. 2 (a)), the lateral surface of the nanowire is represented mainly by the (111) faces with the minimum surface energy density, $\sigma$. On the (100) faces, $\sigma$ is also relatively low. The arising instability is followed by the formation of necking regions with slopes that are formed by facets with greater values of $\sigma$. Thus, the average value of the parameter $\beta$ in Eq. (9) is positive, i.e. $\beta > 0$, that corresponds to the excitation of long-wave modulations of the nanowire radius: $\lambda_{max}/r_0 > 9$. The results of numerical experiments (Fig. 3) confirm this preliminary estimate. The data presented in Fig. 3 were obtained by blocking the evaporation of atoms from the surface of a nanowire ($P_{filter}$=0). That is, the development of instability is associated only with surface diffusion, which was assumed in the derivation of Eq. (9).

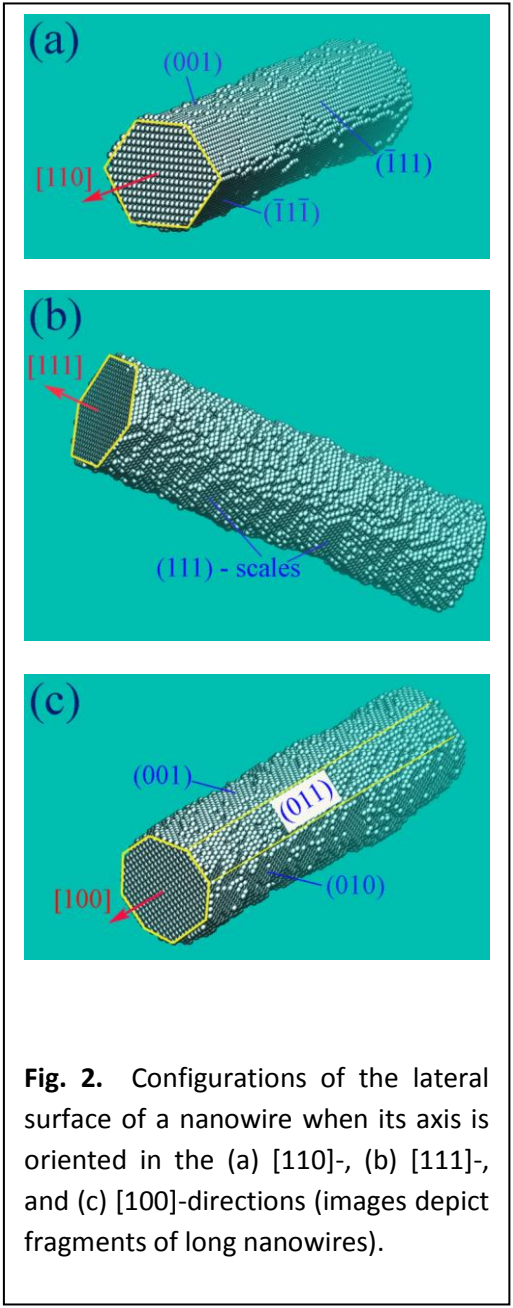

**Fig. 2.** Configurations of the lateral surface of a nanowire when its axis is oriented in the (a) [110]-, (b) [111]-, and (c) [100]-directions (images depict fragments of long nanowires).

One can see that the breakup parameter, $\Lambda/r_0$, monotonically decreases with increasing temperature. According to the results presented in Fig. 3, the ratio, $\Lambda/r_0$, is close to 9 at high temperatures (Fig. 3C; $\Lambda/r_0 \approx 10$) and significantly exceeds this value in the cold regime (Fig. 3A; $\Lambda/r_0 \approx 14.5$). The physical interpretation of this experimentally observed fact **[26]** is straightforward in our numerical model. The jump frequency of bound atoms - that is locally determined by the value of $p_{jump} = p^{m_0} = \exp(-m_0 \times \Delta/kT)$ (see Section 2) - is inhomogeneous along a nanowire surface ($p_{jump}$ is minimal at widening regions). At low temperatures, this inhomogeneity becomes more pronounced. Naturally, the higher surface activity of atoms in necking regions leads to the accumulation of atoms in front of the zones of widening and results in increasing the wavelength of surface perturbations.

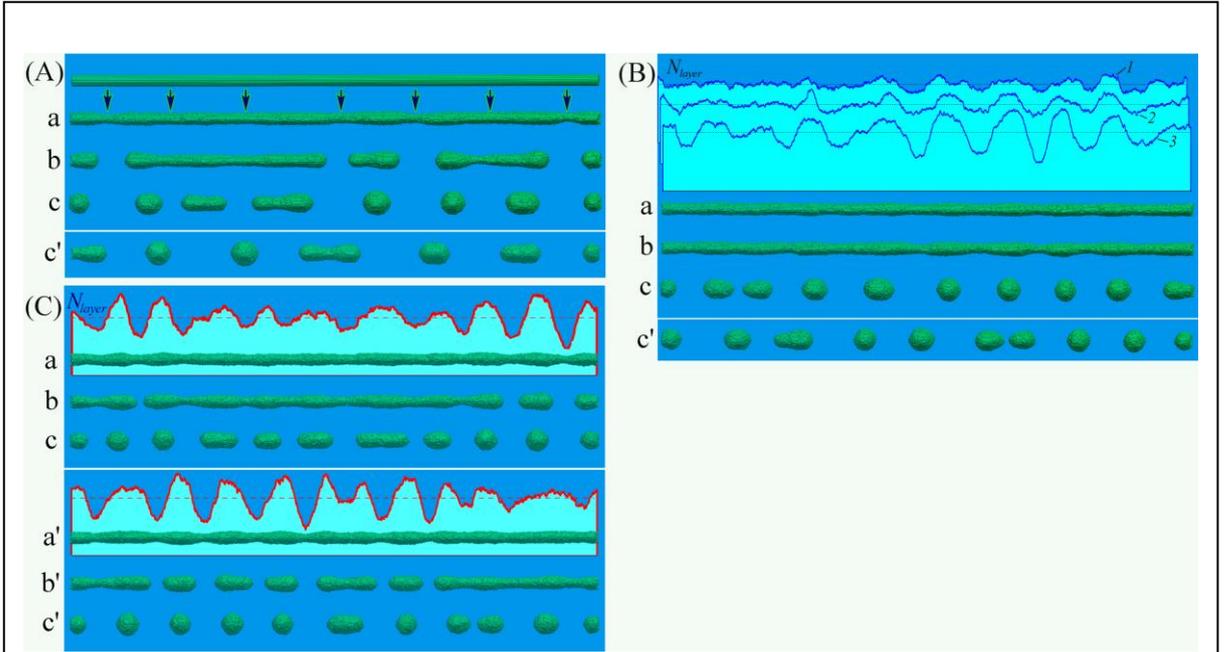

**Fig. 3. Disintegration of a nanowire with its axis along the [110]-orientation when vaporation is blocked; $L = 510$, $d_0 = 10$, the initial number of atoms in the nanowire $N_0 \approx 167400$. (A)** Cold regime: $\alpha = 0.96$ and $p = 0.71$; in subparts (a), (b), and (c): $t = \{7.5, 10.5, 13.5\} (\times 10^6)$ MC steps, respectively, and $\Lambda/r_0 \approx 14.5$. Configuration (c′) is the result of another random MC simulation at $t = 16.7 \times 10^6$ MC steps. **(B)** Warm regime: $\alpha = 0.9$ and $p = 0.725$; in subparts (a), (b), and (c): $t = \{1.5, 3.0, 6.8\} (\times 10^6)$ MC steps, respectively. The upper inset shows the distribution of $\widehat{N}_{layer}$ along the nanowire axis at $t = \{0.75, 1.5, 3.0\} (\times 10^6)$. $\widehat{N}_{layer}$ is the number of atoms in the atomic layers that are perpendicular to the axis of the nanowire, relative to the corresponding average value along the nanowire. Configuration (c′) is the result of another random MC simulation at $t = 7.74 \times 10^6$ MC steps and $\Lambda/r_0 \approx 11.3$. **(C)** Hot regime: $\alpha = 0.8$ and $p = 0.752$; in subparts (a), (b), and (c): $t = \{1.2, 1.8, 2.4\} (\times 10^6)$ MC steps, respectively. Configurations (a′), (b′), and (c′) show the results of another set of random MC simulations at $t = \{1.5, 2.1, 3.45\} (\times 10^6)$, where $\lambda_{max}/r_0 \approx 9.7$ and $\Lambda/r_0 \approx 10$. In this figure and in the sequel, upper insets present $\widehat{N}_{layer}$ – distributions for (a)-configurations.

The results presented in Fig. 3 show that under long-wave perturbations, the number of nanodroplets after the breakup of the nanowire coincides with the number of modulation periods of its cross section, i.e. $\Lambda = \lambda_{max}$. Such a coincidence is violated in the case of [100]-orientation (see below), when, at the stage of the breakup of the nanowire, the developed shorter wavelength perturbations merge into separate nanoclusters, which results in inequality $\Lambda > \lambda_{max}$.

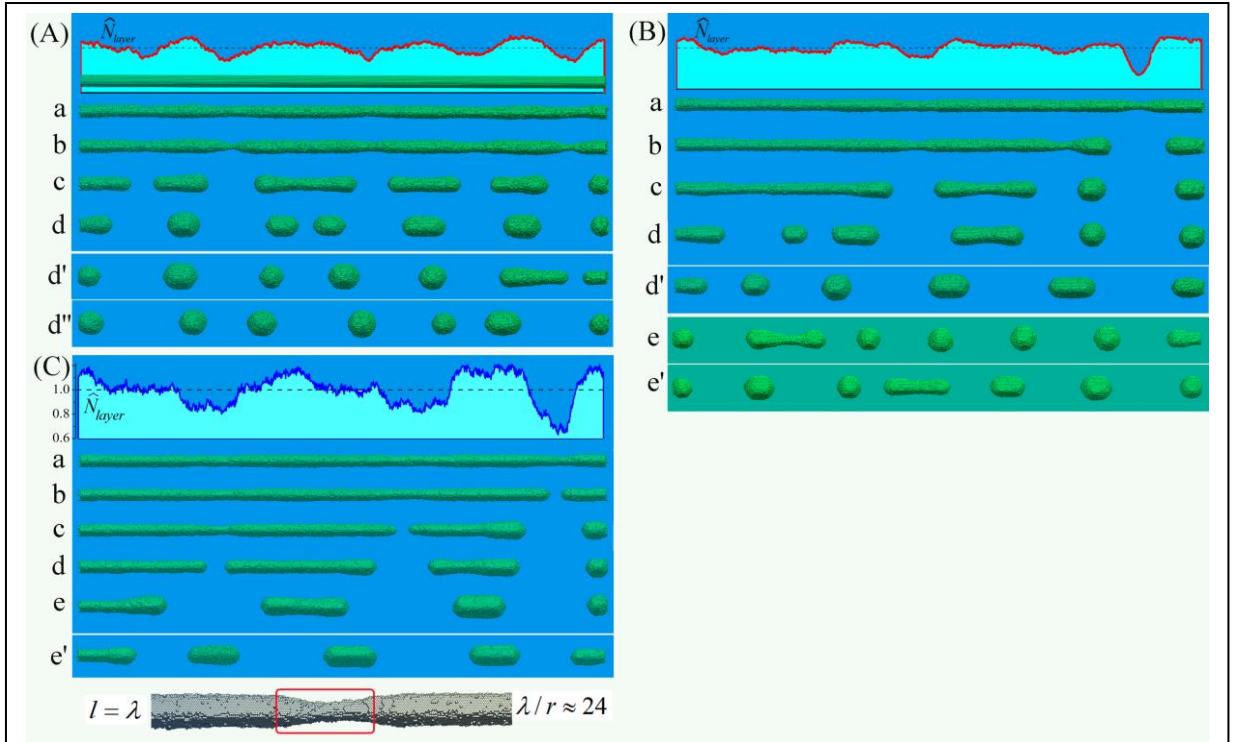

**Fig. 4**. **Disintegration of a nanowire with its axis along the [110]-orientation ($L = 510$). The effect of exchange by atoms between the surface and the near-surface layer is taken into account ($P_{filter} = 1$).** (**A**) Hot regime: $\alpha = 0.8$ and $p = 0.752$; $d_0 = 10.9$, $N_0 \approx 182.5 \times 10^3$, $N_t \approx 168 \times 10^3$, and $d_{eff} \approx 10.4$; (a) to (d): $t = \{2.1, 2.83, 3.7, 4.75\}(\times 10^6)$ MC steps. Configurations (d) and (d′) are the results of two other random MC simulations at $t = 5 \times 10^6$ and $t = 6.35 \times 10^6$, respectively ($\Lambda/r_{eff} \approx 16$). (**B**) Warm regime: $\alpha = 0.9$ and $p = 0.725$, $d_0 = 10.2$; $N_0 \approx 164.2 \times 10^3$, $N_t \approx 157.5 \times 10^3$, and $d_{eff} \approx 10$; (a) to (d): $t = \{5.5, 8.0, 10.0, 13.3\}(\times 10^6)$ MC steps. Configuration (d′) is the result of another random MC simulation at $t = 11.5 \times 10^6$ MC steps ($\Lambda/r_{eff} \approx 20$). Configurations (e) and (e′) are the results of two other random MC simulations when the evaporation is reduced ($P_{filter} = 0.3$); (e): $t = 16.5 \times 10^6$; (e′): $t = 14.1 \times 10^6$ MC steps. (**C**) Cold regime: $\alpha = 0.96$ and $p = 0.71$; $d_0 = 10.25$, $N_0 \approx 173100$, $N_t \approx 167300$, and $d_{eff} = 10.1$; (a) to (e): $t = \{12.0, 14.4, 17.4, 18.9, 24.5\}(\times 10^6)$ MC Steps. Configuration (e′) is the result of another random MC simulation at $t = 21 \times 10^6$ MC steps ($\Lambda/r_{eff} \approx 24$). The lower inset shows a fragment of the nanowire at time $t = 10.5 \times 10^6$ MC steps.

Note that significant variations in the breakup parameter, $\Lambda/r_0$, as a function of temperature are observed in previously reported experiments with Cu **[26]** when the temperature increases from 400 °C ($\Lambda/r_0 \approx 13 - 15$) to 500 °C. In the second case, the diameter of the formed nanodroplets is such that the gap between them is approximately equal to their diameter. For this geometric ratio the breakup parameter is: $\Lambda/r_0 \approx 9 - 10$. However, in experiments with Au reported in [31] and [28], the observed ratio $\Lambda/r_0$ is up to $20 - 24$ (at $T \approx 500\,°C$) and demonstrates a record high value $\Lambda/r_0 \approx 30 - 35$ at $T \approx 400\,°C$. We obtain similar results if the evaporation of bound atoms from the nanowire surface in taken into account ($P_{filter}=1$) (see Fig. 4). The effect of atom exchange between the nanowire surface and the surface layer of free atoms increases the parameter $\Lambda/r_{eff}$ from 10 to 16 in the hot regime and from 14.5 to 24-30 in the cold regime. The effective nanowire radius is

$$r_{eff} = r_0\sqrt{N_t/N_0}, \tag{11}$$

where $N_t$ is the number of atoms in the nanowire after establishing an equilibrium between the surface and the vapor in a short time. This number slightly differs from the initial value, $N_0$, and remains constant at all stages of breakup, changing only due to thermal fluctuations.

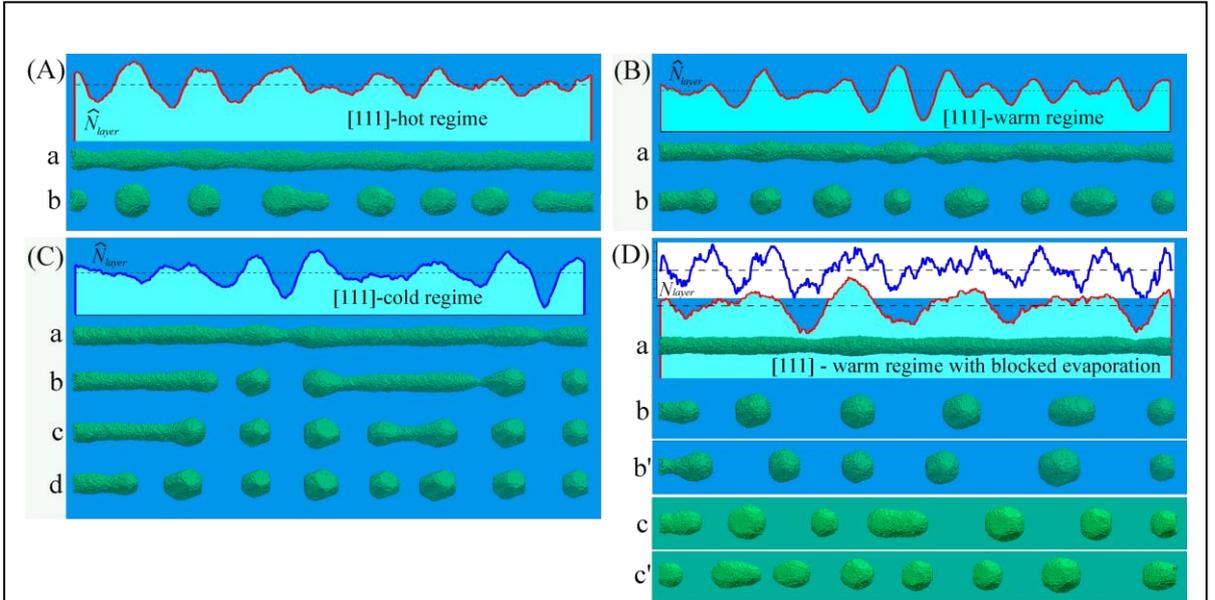

**Fig. 5. Disintegration of a nanowire with its axis along the [111]-orientation ($L = 360$). (A)** Hot regime: $\alpha = 0.8$ and $p = 0.752$; $d_0 = 12.3$, $N_0 \approx 171.2 \times 10^3$, and $N_t \approx 161.5 \times 10^3$; (a): $t = 1.2 \times 10^6$, and (b): $t = 2.25 \times 10^6$ MC steps. **(B)** Warm regime: $\alpha = 0.9$ and $p = 0.725$; $d_0 = 12$, $N_0 \approx 160 \times 10^3$ and $N_t \approx 156.3 \times 10^3$; (a): $t = 2.25 \times 10^6$, and (b): $t = 4.5 \times 10^6$ MC steps. **(C)** Cold regime: $\alpha = 1.2$ and $p = 0.65$; $d_0 = 12$, $N_0 \approx 160 \times 10^3$, and $N_t \approx 159.6 \times 10^3$; (a) to (d): $t = \{24, 30, 36, 45\}(\times 10^6)$ respectively. **(D)** Warm regime with blocked evaporation: $\alpha = 0.9$ and $p = 0.725$; $d_0 = 11.5$ and $N_0 \approx 150 \times 10^3$; (a): $t = 13 \times 10^6$, and (b): $t = 24 \times 10^6$ MC steps ($\Lambda/r_0 \approx 12.5$). The distributions of $\widehat{N}_{layer}$ along the nanowire are shown at times $t = 4.5 \times 10^6$ (the blue curve; $0.85 \leq \widehat{N}_{layer} \leq 1.15$) and $t = 13 \times 10^6$ MC steps. Configuration (b') is the result of another random MC simulation at $t = 27 \times 10^6$ MC steps. Configurations (c) and (c') are the results of two other random MC simulations when the evaporation is reduced ($P_{filter} = 0.3$); (c): $t = 14.7 \times 10^6$, and (c'): $t = 10.2 \times 10^6$ MC steps.

A significant increase in the $\Lambda/r_{eff}$ parameter (compare the results of Fig. 3 and Fig. 4) and the sensitivity of this parameter to temperature changes are inherent to only the [110]-orientation. The breakup characteristics of a nanowire with an [111]-orientation are completely opposite. If the evaporation of atoms from the surface of the nanowire is blocked ($P_{filter} = 0$), then in the warm mode $\Lambda/r_0 \approx 12.5$ (see Fig. 5D). Accounting for detachment-reattachment processes ($P_{filter} = 1$) does not increase (as it occurs in the [110]-orientation), but reduces the breakup parameter to the classical $\Lambda/r_{eff} \approx 9$ regardless of temperature (see Fig. 5A, B, and C).

The reasons for this striking difference can be explained on the basis of the structural features of the side surface of the nanowire in the cases under consideration (in the [110] and [111]-orientations), as discussed below.

The flux density of atomic evaporation from the nanowire surface, $\Phi^{(-)}$, is a local characteristic that, in a rough approximation, depends on the average surface curvature. The deposition of free atoms (the density of their reverse flow, $\Phi^{(+)}$) is associated with the distribution of surface curvature along the entire nanowire and is its integral result. The balance of these flows ($\Phi^{(+)}$ and $\Phi^{(-)}$) is inhomogeneous along the nanowire surface (a detailed analysis of this balance was carried out in our work **[47]** for Si; see the Supplementary Materials). For the sinusoidal perturbations of the wire radius, $\varepsilon \ll r_0$, in Eq. (4), the value $\Phi^{(+)}$ in the narrowing regions is less than in the widening regions if the wavelength is $\lambda \lesssim 10 r_0$. In this case, conditions arise for the development of instability, which is similar to the unstable growth of the shell during the diffusion deposition of material on the nanowire **[23]** (this type of instabilities may be qualitatively represented by the model developed by Mullins and Sekerka **[49]**). It is this mechanism that manifests itself in the dynamics of the nanowire in the [111] orientation. A very important factor is that the surface of the nanowire before the breakup has a scaly structure in the widening zones (see Fig. 2B) composed of fragments of [111] faces (a result of thermal roughening **[50]**). Free atoms deposited on such a surface are captured by numerous corner vacancies, which contribute to the further redistribution of atoms to widening regions and the development of instability (see Fig. 5A, B, and C). Turning off this mechanism of atomic transfer in the regime of blocked evaporation ($P_{filter} = 0$) leads to an expected result that may be obtained on the basis of Eq. (9).

Note that the [111] scaly lateral surface (Fig. 2B) corresponds to a rather low surface energy density. At the same time, the slopes of constrictions arising at the developed stage of instability are formed by the fragments of [111] and [100] faces **[39]** with a larger value of $\sigma$, which corresponds to $\beta > 0$ in Eq. (9) and, therefore, $\lambda_{max} > 9$. Such a qualitative prediction is in agreement with the data of Fig. 5D: $\Lambda/r_0 \approx 12.5$.

Returning to the analysis of [110]-nanowires, note that, in contradiction to [111]-nanowires, the effect of exchange by atoms between the surface and the near-surface layer leads to significant increase in the parameter $\Lambda/r_{eff}$ (see Fig. 4). In this case, the side surface of the nanowire is mainly composed of [111] facets (see Fig. 2(a) and the lower inset in Fig. 4C)). Atoms deposited on the mainly flat facets [111] have only three bonds and a high coefficient of surface diffusion. If the length between neighbor necking regions is not sufficient, then these deposited atoms will not have enough time to form new clusters on the side facets. As a result, the surface diffusion flows of bonded atoms directed to the necking regions from the left and right (see the lower inset in Fig. 4C, red square) can lead to its shortening, which diminishes the intensity of the detachment process at this region. Thus, the instability may arise only for more elongated necking and widening regions, than in the case of blocked evaporation.

Finally, we analyze the breakup of a [100]-nanowire. Results presented in Fig. 6 show that there is a weak reaction of the breakup parameter, $\Lambda/r_{eff}$, on temperature variations and the effects of surface-vapor exchange in this case. The structure of the lateral nanowire surface before breakup is shown in. Fig. 2(c). It is composed of four [100]- and four roughened [110]- facets. Taking into account that the slopes of the necking regions are mainly formed by flat [111] facets **[39]** with the lowest surface energy density, we may claim that the value $\beta < 0$ in Eq. (9), and the length of the excited perturbations must be less than $9r_0$.

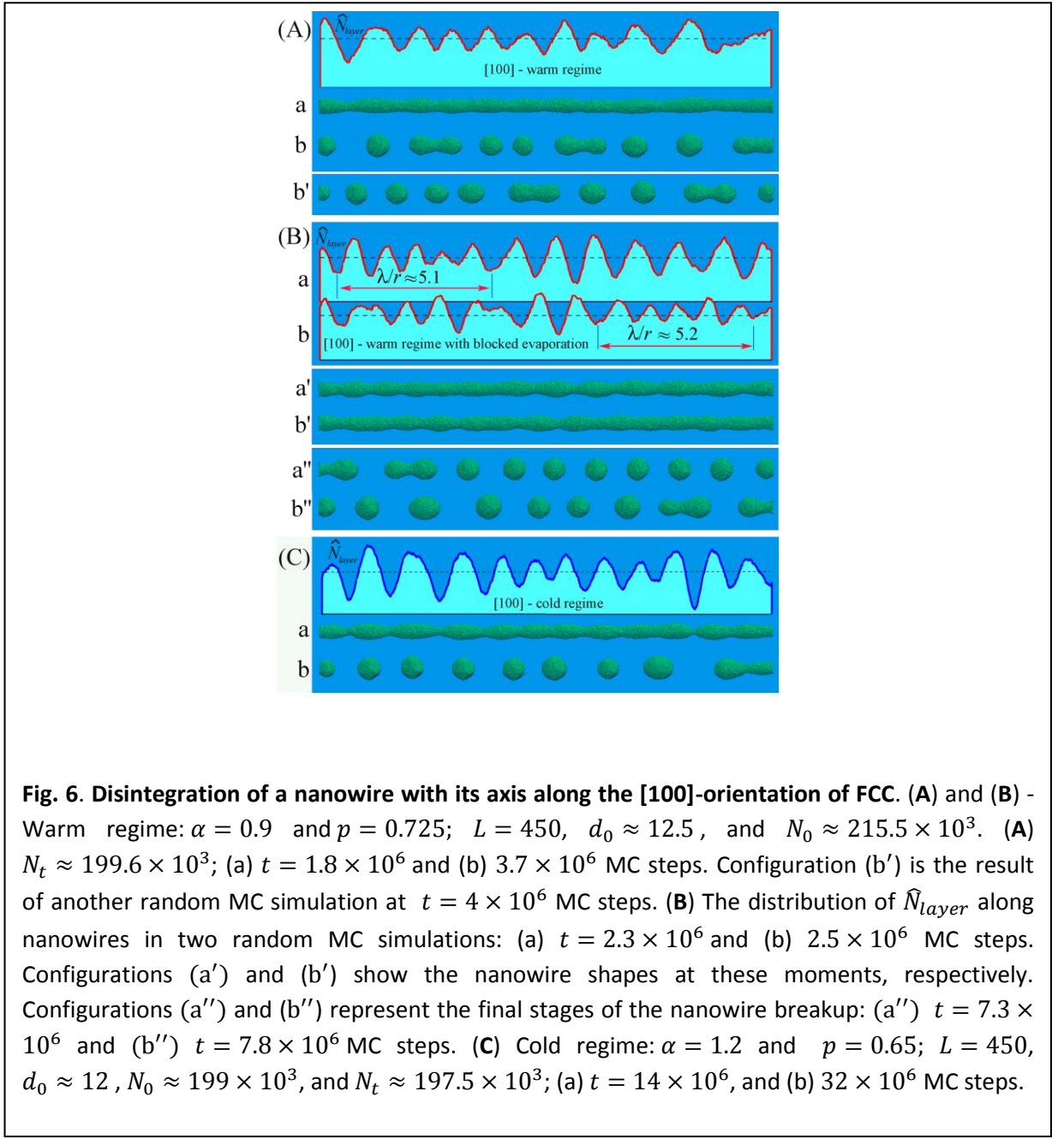

**Fig. 6**. **Disintegration of a nanowire with its axis along the [100]-orientation of FCC**. (**A**) and (**B**) - Warm regime: $\alpha = 0.9$ and $p = 0.725$; $L = 450$, $d_0 \approx 12.5$, and $N_0 \approx 215.5 \times 10^3$. (**A**) $N_t \approx 199.6 \times 10^3$; (a) $t = 1.8 \times 10^6$ and (b) $3.7 \times 10^6$ MC steps. Configuration (b′) is the result of another random MC simulation at $t = 4 \times 10^6$ MC steps. (**B**) The distribution of $\widehat{N}_{layer}$ along nanowires in two random MC simulations: (a) $t = 2.3 \times 10^6$ and (b) $2.5 \times 10^6$ MC steps. Configurations (a′) and (b′) show the nanowire shapes at these moments, respectively. Configurations (a″) and (b″) represent the final stages of the nanowire breakup: (a″) $t = 7.3 \times 10^6$ and (b″) $t = 7.8 \times 10^6$ MC steps. (**C**) Cold regime: $\alpha = 1.2$ and $p = 0.65$; $L = 450$, $d_0 \approx 12$, $N_0 \approx 199 \times 10^3$, and $N_t \approx 197.5 \times 10^3$; (a) $t = 14 \times 10^6$, and (b) $32 \times 10^6$ MC steps.

A numerical estimate of $\lambda_{max}$ in this case can be obtained on the basis of the following assumptions. Since the lateral surface of the nanowire is bounded by (100)- and scaly (110)-facets, the value of $\sigma_0$ in Eq. (5) can be set equal to $\sigma_{(100)}$. The angle, $\gamma$, between the normal to the axis of the nanowire and the normal to the cutting planes (111), which form the slopes of the necking regions, is equal to $\gamma = 0.62\ rad$. Then

$$\beta = \frac{1}{2}\frac{d^2}{d\gamma^2}\left(\frac{\sigma}{\sigma_0}\right) \approx \frac{\sigma_{(111)}/\sigma_{(100)} - 1}{\gamma^2} \approx -0.33, \qquad (12)$$

which leads to the ratio $\frac{\lambda_{max}}{r_0} = 9\sqrt{1 + 2\beta} \approx 5.2$ .

This prediction is in agreement with results shown in Fig. 6B at some regions of the nanowire where the deviations $\Delta\widehat{N} = \widehat{N} - 1$ may be associated with the linear stage of the surface dynamics. Moreover, the similar shortwave, $\lambda_{max}/r_0 \approx 5.7$, modulations, $\varepsilon$, of the nanowire radius ($\varepsilon \approx 0.5 r_0$) were observed in **[32]**. In the final stage of instability (see Fig. 6B), when the nanowire radius is rather visible, $\lambda_{max} \approx 6.3$ according to $\widehat{N}(x)$ − Fourier transform. However, these short-wave radius modulations cannot result in single droplets of the corresponding size at the nonlinear stage of breakup, so they either merge into larger-scale clusters (in all the cases under consideration: $\Lambda/r_{eff} \approx 8.3 - 9.5$) or remain in metastable state for a while **[32]**.

Now, we briefly discuss the reasons for the insensitivity of a [100]-nanowire breakup parameter, $\Lambda/r_{eff}$, to temperature changes. For sinusoidal-like short-wave perturbations ($\lambda/r_{eff} \lesssim 8$ and $\varepsilon \lesssim 0.4 r_0$), the flow density $\Phi^{(+)}$ is the largest in the widening regions **[47]**, although the surface curvature in these zones, $\kappa_w$, may be lower than in the narrowing zones, $\kappa_n$. If the sublimation flux density from the nanowire surface, $\Phi^{(-)}$, is associated with the local curvature, $\kappa(x)$, of the surface, then in the narrowing zones $\Phi^{(-)} > \Phi^{(+)}$. Additional atomic transport in the near-surface layer accelerates the breakup processes. This effect is realized in the case of [100]-orientation with the maximal order of rotational symmetry and the sinusoidal-like surface perturbations (see the data presented in Figs. 6A and B).

The slight increase in the wavelength, $\lambda_{max}$, in Fig. 6A, compared to Fig. 6B, is a result of the thermal roughening of the nanowire surface caused by its exchange with the surrounding vapor of free atoms, which decreases the level of the anisotropy of $\sigma$. On decreasing the temperature, this exchange is suppressed, and $\lambda_{max}$ takes the previous value (see Figs. 6B and C).

## 5. Discussion and conclusion.

The main conclusion that can be made on the basis of the above results is as follows. Physical factors that determine variety scenarios of the breakup of nanowires depending on temperature, radius, type of material can be qualitatively but rather unambiguously interpreted on the basis of a simple Monte Carlo model containing two dimensionless thermodynamic parameters. For example, the results of the experiments **[29]** are predictable as follows: for the given radius, $r_0$, the breakup parameter, $\Lambda$, of a poly-crystalline nanowire is much shorter than for the single-crystalline nanowire with [110] orientation, since in the first case the anisotropy of the surface energy density is significantly suppressed. The results of **[51]** are also clear - Au-based alloyed nanowires (AuCu, AuPd, and AuPt) demonstrate high thermal stability. An increase in $\varepsilon$ and $\Delta$ (the pair binding free-energy and the activation free-energy barrier, respectively) is equivalent to a sharp decrease in the effective temperature of the nanowire.

As stated above, the exchange by free atoms between the nanowire surface and the near-surface layer may play an ambiguous role. Qualitative analysis of the relevant results presented in Fig. 3, 5D, 6B (the sublimation of atoms from the nanowire surface is absent) is the following.

The development of instability should be accompanied by a decrease in the free energy of the nanowire. In the case of an isotropic density of surface energy, $\beta = 0$, the area of the lateral surface, $S$, should decrease, i.e. $\Delta S < 0$. With the sinusoidal perturbations of the nanowire radius, this happens only for $\lambda > 2\pi r$. The value $|\Delta S|$ increases with increasing wavelength and this is reflected in Eqs. (8a) and (8b) by the factor $[1 - \hat{k}^2]$ on its right-hand side. However, it should be noted that with increasing wavelength, the mass of the transported substance also increases, which is reflected by the factor $1/\hat{k}^2$ on the left side of Eqs. (8a) and (8b). The optimal ratio between the mentioned values, corresponding to the maximum instability increment, $\gamma$, is achieved for $\lambda \approx 9r$ (the result is well known in theory **[24, 25]**). For $\beta \neq 0$, two options are realized.

If the surface energy density, $\sigma$, decreases on the slopes of necking regions with an increase in the angle of inclination ($\beta < 0$ in Eq. (5)), then for the same amplitude $\varepsilon(t)$, the effect of an additional decrease in surface energy is more pronounced for shorter wavelength disturbances, which is a result of the $\sigma$-anisotropy. The maximum of the increment, $\gamma$, shifts toward shorter waves, $\lambda < 9r$, and the value of $\gamma$ itself increases due to the corresponding decrease in the mass of matter transported. Note that for a large value $|\beta|$, the optimal ratio between the change in free energy and the mass transferred (see Eq. (3)) can be realized for very short wavelengths, when changes in the area of the lateral surface of the nanowire are minimal. However, the excitation of such short waves at the linear stage of instability can lead to the phenomenon that at the nonlinear stage, neighboring proto-droplets will merge into larger clusters. The described effects are demonstrated by the results presented in Fig. 6. Noticeable modulations of the number of atoms in atomic layers in Fig. 6B are developed with a wavelength of $\lambda \approx 6.3r$, which is substantially less than $9r$. At the nonlinear stage of breakup, the number of drops in all variants of Fig. 6 is such that $\Lambda \approx 9.3r$. Thus, the value of the breakup parameter, $\Lambda/r$, determined in experiments, hides the real dynamics of the process associated with the pronounced anisotropy ($\beta < 0$) of $\sigma$.

Completely different decay scenarios are realized in cases when $\beta > 0$ (in the [111]- and [110]-orientations). The most expressive are the features of the breakup of a nanowire with a [110]-orientation. As noted above, its lateral surface is characterized by a minimum average surface energy density (see Fig. 2A). On the surface of the arising necks, the value of σ is higher and in all the presented temperature regimes (see Fig. 3) $9r < \lambda_{max} \approx \Lambda$ (merging of neighboring drops practically does not occur). An increase in these decay parameters with decreasing temperature (up to $\Lambda/r \approx 14.5$) is associated with the inhomogeneous surface mobility of atoms, which leads to a noticeable deviation of the developing perturbations from the sinusoidal shape and a pronounced localization of the breakup regions. This feature of the [110]-orientation was observed in the results of experiments with copper **[26]**.

Depending on the material, diameter, and temperature, the formation of a shell of free atoms around a nanowire can result in significant modifications to the dynamics of its decay. The nature of these modifications depends on the intensity of surface diffusion in the narrowing and broadening regions, the sublimation intensity in these regions, and the distribution of diffusion fluxes of free atoms from the surface layer onto the nanowire surface.

Obviously, the dominance of sublimation fluxes, $\Phi^{(-)}$, in the narrowing regions over the reverse flow of free atoms, $\Phi^{(+)}$, contributes to the development of instability that is caused by surface diffusion. However, the realization of such a ratio with the corresponding self-tuning of the surface can lead to a substantial rearrangement of the spectrum of excited perturbations, as the results of effects that are analogous to the roughening transition **[50]**. For sinusoidal radius modulations, a dominance of the back-flows, $\Phi^{(+)}$, at the widening regions (as compared to that at the necking regions) can be realized either for short-wave perturbations ($\lambda_{max} < 8r$ and the shadow effect in diffusion becomes apparent) or for longer perturbations ($\lambda_{max} \gtrsim 11r$, when $\Phi^{(+)} \sim 1/r(x)$).

The first option is realized in the case of nanowires with [111] orientation, when the slow surface diffusion along the scaly-surface structure does not suppress the growth of short-scale bulges, and the breakup parameter $\Lambda/r_{eff}$ reduces from 12.5 (see Fig. 5B) to the classical value of 9, independently of the temperature regime (see Figs. 5A and B). Thus, the effects of the exchange by atoms between the surface of a nanowire and the enveloping near-surface layer hide the manifestation of anisotropy $\sigma$ in real experiments.

The second option is clearly manifested in the nanowires with an [110] orientation, the lateral surface of which is mainly formed by (111) faces. The relatively high diffusion rate on such planes inhibits the formation of nanoclusters (new layers in the widening regions) on them. Free atoms settling on these faces drift to the narrowing regions and inhibit their extension along the nanowire. Instabilities arise only with increased distances between necks, which is demonstrated by the results presented in Fig. 4. Note that the effects of sublimation noticeably increases the breakup parameters of nanowires even at a relatively low intensity (see Fig. 4B, configurations (e) and (e$'$); $P_{filter} = 0.3$).

Hence, the manifestation of the anisotropy effects of the physical properties of the surface can be clearly demonstrated by a significant deviation of the breakup parameter from its classical value, 9. However, the absence of the noticeable deviation from this value, 9, can also be a result of the complex interaction of various anisotropy factors at different stages of breakup.